Critique of arXiv submission 2308.15623, "Discovery of Spherules of Likely Extrasolar Composition in the Pacific Ocean Site of the CNEOS 2014-01-08 (IM1) Bolide" by A. Loeb et al.

Steve Desch[*] and Alan Jackson[**]

[*] Arizona State University, School of Earth and Space Exploration, PO Box 876004, Tempe, Arizona, 85287-6004
[**]Towson University, Department of Physics, Astronomy and Geosciences, 8000 York Rd., Towson, Maryland, 21252

**Abstract**

Recently a manuscript by Loeb *et al.* was uploaded to arXiv[1] (preprint 2308.15623) that asserted that the CNEOS bolide 2014-01-08 was interstellar; that spherules recovered from the seafloor near the airburst were associated with this bolide; that they had Fe isotopic ratios indicating origin as micrometeorites; that they had unusual chemical compositions enriched in Be, La and U, never seen before in micrometeorite spherules; that these compositions were formed in the magma ocean stage of a differentiated extrasolar planet; and that the Be abundance reflected passage through the interstellar medium. Despite not being peer-reviewed, this uploaded manuscript has been reported by media outlets as "published", and its conclusions have been widely distributed as fact. The purpose of this manuscript is to provide potential peer reviewers and the general public with an appreciation of the multiple fatal flaws with the manuscript's arguments. We discuss the published evidence that the 2014-01-08 bolide is not interstellar. We show that there is no statistical spatial correlation of a chemical signature or even number of recovered spherules with the 2014-01-08 bolide. We demonstrate that the Fe isotopic ratios decisively indicate an origin in our Solar System, with > 99.995% probability. We demonstrate that the unusual enrichments in La, U, etc., have in fact been observed in micrometeorites before and attributed to terrestrial contamination; and that the Be abundances are similarly consistent with those of ferromanganese nodules, after reacting with sea water. Far from being exotic particles from an extrasolar planet, the spherules collected and analyzed by Loeb *et al.* appear to be just like those found around the world, with a Solar System origin and compositions modified by tens of thousands of years residence at the ocean bottom.

---

[1] https://arxiv.org/pdf/2308.15623.pdf

**I. Introduction**

On August 29, 2023, Loeb *et al.* uploaded a manuscript, "Discovery of Spherules of Likely Extrasolar Composition in the Pacific Ocean Site of the CNEOS 2014-01-08 (IM1) Bolide" to the arXiv preprint server[2]. In it, the authors reported on their recovery of spherules from seafloor near Papua New Guinea, and their subsequent analysis. This manuscript claimed that these spherules were likely of interstellar origin, the first such samples ever recovered. This claim, if verified, would represent a significant scientific advance, allowing the first laboratory analysis of macroscopic quantities of material from another solar system, enabling direct tests of theories about how stars and planets are born in our Galaxy.

Because of the boldness of the claim, it might be expected that Loeb *et al.* exercised due diligence to eliminate alternative, more-prosaic explanations for the data, and would report their results carefully and deliberately. They did not. Loeb *et al.* hardly considered, and did not test, the simplest alternative hypothesis: the spherules they found were of a common and well-understood type found worldwide, from Solar System asteroids. Apparently, the authors did not seek expert opinion, and posted the "discovery" manuscript before it had been peer-reviewed and accepted, or even submitted to a science journal. The Galileo Project headed by Loeb did, however, put out a press release the same day touting the manuscript[3], picked up by national media such as *USA Today*[4].

While the manuscript focused on the possibility that the spherules are from a natural object, it also suggested they "may reflect an extraterrestrial technological origin." The framing of the *USA Today* article advances this by repeating claims that the spherules are "unmatched to any existing alloys." About the paper, on August 31 Loeb said to *Salon*[5]: "The ideal scenario is that in addition to tiny fragments, we would find a piece of an advanced technological device, like the hundredth version of the iPhone… I would love to press a button on such an object." Not coincidentally, Loeb's book, *Interstellar: The Search for Extraterrestrial Life and Our Future in the Stars*, was published the same day the manuscript and press releases were posted.

The net result is that Loeb *et al.* have affixed in the public consciousness the unsubstantiated—and probably false—idea that they have recovered interstellar spherules; and they have irresponsibly led the public to conflate "interstellar" with "extraterrestrial technology."

Ideally, the peer review process would identify the fatal flaws in their conclusions, finding no evidence for an interstellar origin, the manuscript would not be published in a peer-reviewed journal or given legitimacy, and it would be forgotten. However, by posting the manuscript on the arXiv preprint server before peer review, by issuing premature press releases, and by making very public and unsubstantiated claims, Loeb *et al.* have circumvented the normal peer-review process. This demands a response to their points in an equally public venue.

---

[2] https://arxiv.org/pdf/2308.15623.pdf
[3] https://projects.iq.harvard.edu/galileo/news/spherule-analysis-finds-evidence-extrasolar-composition
[4] https://www.usatoday.com/story/news/nation/2023/08/29/metallic-spheres-interstellar-origin-avi-loeb-finds/70699783007/
[5] https://www.salon.com/2022/08/31/cneos-2014-01-08-loeb-alien-asteroid/

This article represents our attempt to make public the many clear and fatal flaws in the arguments made by Loeb *et al.* Far from being interstellar objects, their own data demand the spherules be from our Solar System. Although the authors do not recognize it, the spherules are of a common type, and evidently have reacted with seawater for tens of thousands of years, explaining their enrichments in Be, La, and U, among other elements. They are probably not even associated with the 2014-01-08 bolide, which was probably not interstellar anyway.

We hope this manuscript will serve as a resource for multiple audiences. For the general public, we hope to communicate in a clear and accessible way that Loeb *et al.* have **not** discovered objects from beyond our Solar System, let alone anything technological. For Loeb and co-authors, we hope to provide a referee report that identifies the major flaws that should be rectified, if that is even possible, before their work should be published. A pillar of the scientific method is self-correction—the refinement or even abandonment of theories that simply don't match observations. However, anticipating that Loeb *et al.* will not abandon their premature and unwarranted claims, we hope that this manuscript will serve as a resource for potential peer reviewers, so that this deeply flawed work does not enter the scientific literature.

Below we summarize the claims made by Loeb *et al.* We provide an overview of the major flaws, then discuss each one in turn.

## II. Summary of the manuscript by Loeb *et al.*, and outline of this article

We begin with a summary of the claims made by Loeb *et al.,* and the logic of their argument. We then summarize the critiques of each point and outline this paper.

*Summary of the claims by Loeb et al*.

The manuscript describes the recovery and analysis of spherules from the ocean floor in the vicinity of where the authors believe the Center for Near-Earth Object Studies (CNEOS) 2014-01-08 bolide[6] disintegrated and deposited material. A magnetic sled was employed, resulting in the collection of about 700 spherules, of which 57 were analyzed for bulk and trace chemistry, including 11 for Fe isotopic compositions. It is argued that a subset of at least 5 of these spherules, with distinctive enrichments in Be, La, U, etc., which they term "BeLaU" spherules, are from the 2014 bolide and are of extrasolar origin. This claim of extrasolar material is based on the following chain of logic.

1) The authors claim that the 2014-01-08 bolide was interstellar in origin, moving too fast to have been bound in orbit around the Sun when it struck the Earth.

---

[6] https://cneos.jpl.nasa.gov/fireballs/

2) The authors assert that significant quantities of material from this bolide would have survived atmospheric entry, to be deposited on the seafloor, mostly as millimeter-sized ablation spherules.

3) The authors claim to know the position where these spherules would be deposited, to within an area 1 km wide by 10 km long, the bolide's assumed path.

4) The authors claim that a statistically significant excess of spherules (twice as many) were collected along the assumed path, compared to two control regions far from the path, implying that a large fraction (about half) of the spherules collected along the assumed path are from the 2014 bolide.

5) The authors measured Fe isotopic ratios in 11 spherules, including 3 "BeLaU" spherules, and find all of them lie along the terrestrial fractionation line within errors (< 1‰), and thus show fractionations consistent with Fe loss by vaporization during entry.

6) One of these "BeLaU" spherules ("S21") was found to be a compound spherule, three spherules stuck together while molten. The authors present a back-of-the-envelope calculation to suggest this is also consistent with formation in a fireball.

7) The authors measured elemental compositions of 57 spherules and found 5 "BeLaU" spherules with distinctive enrichments in Be, La, U, and other elements (and depletions in Mg, Ca, etc.) relative to CI chondrites. The authors assert that this compositional pattern is not found in *any* terrestrial samples or other cosmic spherules. By implication the "BeLaU" spherules would have to be extrasolar.

8) The authors claim the "BeLaU" spherules are associated only with the assumed path of the bolide and are absent from regions far from the assumed path, and therefore associated with the 2014 fireball.

9) The authors argue that the "BeLaU" compositions would be (uniquely?) consistent with formation in the magma ocean of a differentiated (extrasolar?) planet.

10) The authors suggest the high abundance of Be, which is produced by spallation of atoms in meteoroids by Galactic cosmic rays (GCRs), is indicative of GCR irradiation during passage through interstellar space.

To claim an interstellar origin for the spherules they collected, all the links in the chain of logic above would have to withstand scrutiny.

*Summary of the critiques to these claims*

In fact, **not a single link** above withstands scrutiny.

1) The 2014-01-08 bolide probably wasn't interstellar. Uncertainties in the velocities of objects in the CNEOS database are not reported, but these can be estimated separately and they are large. There is a > 0.1% probability the 2014-01-08 bolide is not interstellar, which sounds small, but in a catalog of about a thousand (Solar System) bolides, the odds are high that *one* would be that significantly mismeasured and appear to have interstellar velocities. 2014-01-08 appears to be that one.

2) If interstellar, practically none of the 2014-01-08 bolide would have survived entry. If it were traveling at the speeds that were reported (and necessary to be interstellar), then at least 99.8%, and probably > 99.9999% of it would have vaporized in the atmosphere, leaving insignificant quantities to be deposited on the seafloor.

3) The location of the bolide's path is far more uncertain than Loeb *et al.* realize and also almost certainly miscalculated. The CNEOS database reports two locations for the 2014-01-08 fireball, 55 km apart. Even if we pick one location, there are still large inherent uncertainties in trying to pinpoint the location using sound waves reaching a seismometer on Manus Island some 90 km away.

4) There is no evidence the density of collected spherules varied across their search region. The authors did not collect a statistically significant number of spherules from regions off the assumed path to test this. Even if there were differences, it isn't clear they could be attributed to regional concentrations versus differences in collection efficiencies.

5) The Fe isotopic ratios are a smoking gun for the spherules originating in our Solar System. The fact that all the samples lie within < 1‰ of the terrestrial fractionation line (including all the "BeLaU" spherules they measured) indicates a > 99.995% probability that all the spherules (including the "BeLaU" spherules) are from our Solar System. And while the Fe isotopic ratios may indicate fractionation during atmospheric entry, they appear more consistent with terrestrial contamination.

6) The triple spherule S21 almost certainly did not form in a fireball event like the 2014 bolide. It requires a denser environment such as an impact plume. The back-of-the-envelope calculation meant to demonstrate that compound spherules are consistent with a fireball origin actually is in error by a factor of $10^4$. If it *were* true, compound cosmic spherules would be common, which they are not.

7) Despite the authors' assertions, the distinctive "BeLaU" patterns have been seen before, in cosmic spherules (of Solar System origin) from the Indian Ocean (Rudraswami et al. 2016) and ancient micrometeorites from Antarctica (van Ginneken et al. 2021). They are not even dissimilar to (anthropogenic) coal ash (Gallardo 2023).

8) The authors collected too few spherules off the assumed path of the bolide to make any statistical arguments about the association of "BeLaU" spherules with the path. The fact that they found no "BeLaU" spherules off the path is meaningless because they measured only 10 and would expect to find only 1.

9) The "BeLaU" pattern seen in other cosmic spherules has been attributed to terrestrial contamination (Rudraswami et al. 2016; van Ginneken et al. 2021). Few details were provided of how this pattern would arise in an extrasolar magma ocean, but this interpretation has flaws. Contamination by sea water is likely.

10) Finally, the Be abundance seen in these spherules is far too high to be created by cosmic-ray spallation, but is very similar to that seen in ferromanganese nodules, and therefore also consistent with reaction with seawater over tens of thousands of years.

*Outline*

Loeb *et al.* have presented a hypothesis that many of the spherules they've located are interstellar in origin. For this to be tenable, each of the claims above would have to resist falsification, and the chain of logic withstand scrutiny. One might expect a few links above to be weak or hard to verify. Instead, upon minimal investigation, *every single link* in the chain is unsubstantiated. In many cases (points 5, 6, and 7 especially), their own data *negate* their conclusions.

In the sections below, we investigate each of these points one-by-one. We then close with conclusions about how to properly search for interstellar micrometeorites.

**III. Was the 2014-01-08 bolide interstellar?**

The recovery of interstellar materials is, of course, predicated on the 2014-01-08 bolide itself originating outside the Solar System. The manuscript by Loeb *et al.* asserts this and repeatedly assumes that it is, supporting this with the statement: "In 2022 the US Space Command issued a formal letter to NASA certifying a 99.999% likelihood that the [2014-01-08] object was interstellar in origin."

This is not quite true. US Space Command never certified the likelihood was 99.999%. The March 1, 2022 memo from Lt. Gen. John Shaw of Space Force states[7]: "Dr. [sic] Amir Siraj and Dr. Abraham Loeb… authored a paper… [that] reported the meteor as originating from an unbound hyperbolic orbit… with 99.999% confidence," and continued, "Dr. Mozer confirmed that the velocity estimate reported to NASA is sufficiently accurate to indicate an interstellar trajectory." But US Space Command never stated a confidence level for their trajectory calculation, and they did not confirm the asserted 99.999% probability.

---

[7] https://lweb.cfa.harvard.edu/~loeb/DoD.pdf

The number 99.999% does not appear in the refereed papers by Siraj & Loeb (2022a, b). All citations are to its first mention, in the unrefereed work by Siraj & Loeb (2019, arXiv 1904.07224)[8]. There, the probability was estimated assuming relatively small measurement uncertainties in each of the velocity components. This underestimates, by orders of magnitude, the probability the bolide 2014-01-08 originated in our Solar System.

This probability is entirely dependent on the measurement uncertainties in the velocities. The geocentric velocities reported by the Department of Defense to the Jet Propulsion Laboratory (JPL) Center for Near-Earth Object Studies (CNEOS) fireballs database for the 2014-01-08 bolide are[9]: $V_x$=-3.4 km/s, $V_y$=-43.5 km/s, and $V_z$=-10.2 km/s, for a total speed relative to Earth of 44.8 km/s. Considering the time of impact ($17^h05^m34^s$, Jan. 8, 2014) and the rotation of the geocentric frame, this translates to a heliocentric speed of 61.2 km/s at the time of impact. Heliocentric speeds > 42.5 km/s, like this one, indicate an object is not gravitationally bound to the Sun and must be interstellar. It is straightforward to show, though, that if the $V_y$ velocity were mismeasured and really -19.0 km/s, then the speed relative to Earth would be 22 km/s, and the heliocentric speed would be 42 km/s, meaning it would have been in an orbit bound to the Sun. The question of whether the 2014-01-08 bolide was interstellar rests entirely on how unlikely it is for its $V_y$ velocity to have been mismeasured (or misreported) by 24.5 km/s.

Such a large error *sounds* too large to be made by US Government sensors, but what exactly are the errors in the reported velocities? Siraj & Loeb (2019, arXiv 1904.07224) assumed the measurement errors were only ±10% (i.e., ±4.4 km/s uncertainty in $V_y$), based on an unverifiable personal communication (i.e., someone's guess). In other words, they accept that the velocities are imprecise, but at a level equivalent to the determination of an interstellar trajectory for 2014-01-08 being a 5.7σ result (24.5/4.35≈5.7), i.e., only a 0.001% chance of it being from our Solar System.

Already in 2019, though, Brown et al. (2016) and Devillepoix et al. (2019) had investigated the accuracy of the velocities reported in the CNEOS database, by checking them against precise velocities measured for 10 meteors also observed by ground-based camera networks. Only 4 of the 10 meteors' orbits could be accurately reproduced from CNEOS data. Two meteors had radiants (directions of travel) fully 90° off from what was reported in the CNEOS database. Two had speeds off by significant amounts: Buzzard Coulee was reported at 12.9 km/s but really traveled at 18.1 km/s (Milley et al. 2010); and a Romanian bolide was reported at 35.7 km/s but really traveled at 27.8 km/s (Borovicka et al. 2017). Those works suggested, and careful analysis by Brown & Borovicka (2023) and Hajdukova et al. (2023, submitted to *Nature Astronomy*) subsequently confirmed, that the 1σ uncertainties in velocity measurements in the CNEOS catalog are > 8 km/s for fast meteors like the 2014-01-08 bolide.

---

[8] https://arxiv.org/pdf/1904.07224.pdf
[9] https://cneos.jpl.nasa.gov/fireballs/

Therefore, instead of it being a > 5σ detection, the interstellar nature of 2014-01-08 is more like a 3σ detection (24.5/8). Instead of a 0.001% chance it is from our Solar System, there is a 0.1% chance that it is. That is, there is a one-in-1000 chance the Vy velocity has been mismeasured so badly (-43.5 km/s instead of -19.0 km/s) that this meteor could have been in a Sun-bound orbit yet appeared as "obviously" interstellar as 2014-01-08. In isolation, the 2014-01-08 bolide clearly *appears* interstellar. But the CNEOS catalog contains 966 fireballs. If all are from our Solar System but have measurement or reporting errors of 8 km/s, then we should expect (with 66% probability) at least *one* to have been so badly mismeasured it would appear just as clearly interstellar as 2014-01-08. All the other potentially interstellar candidates in the database are much less convincing, and 2014-01-08 happens to be that one fluke.

The situation is exactly analogous to the probabilities of rolling five dice and getting a Yahtzee (all the same number). On any single roll, the odds are 1/1296, or 0.08%. But if you roll the dice 966 times, there is a 52% probability that *one* of those rolls is a Yahtzee. Let's say you happened to roll five sixes on the 610$^{th}$ roll. Should you be accused of cheating and of turning the dice over by hand? After all, the odds of that 610$^{th}$ roll being a Yahtzee were only 0.08%!

It should be noted that a mismeasured velocity, yielding a speed relative to Earth of 22 km/s from a Sun-bound orbit, would imply a lower and more common material strength ~50 MPa (see below), and would be more consistent with the light curve of the 2014-01-08 bolide, as an object at 45 km/s would have been visible by satellites at a higher altitude (Brown & Borovicka 2023). The 2014-01-08 bolide appears more consistent with an origin in our Solar System.

In a catalog with a thousand objects from our Solar System, there was bound to be at least one that was mismeasured (or misreported) and appeared interstellar at the 99.9% level. Loeb *et al.* have cherry-picked that one object and assigned it near-certainty of being interstellar. They choose to dismiss the analyses of Brown et al. (2016), Devillepoix et al. (2019), and Brown & Borovicka (2023) and ignore the evidence that the velocity uncertainties in the CNEOS database are simply larger than the estimate they used. When Loeb was questioned about the reliability of CNEOS data, he is quoted by the *New York Times* on July 24, 2023 as saying, "They are responsible for national security, I think they know what they are doing."[10] Appealing to authority instead of examining evidence and checking one's work is an abrogation of a scientist's duties.

> In summary, velocities of bolides are just not reported very accurately in the CNEOS database. It's entirely possible one component of the velocity of the 2014-01-08 bolide was mismeasured or misreported. Its interstellar status is a 3σ detection, or one-in-a-thousand chance of it being from the Solar System but appearing interstellar. But there was bound to be one object like this among the 966 objects in the CNEOS database, and this is that one. The evidence for the 2014-01-08 bolide being interstellar is entirely unconvincing.

---

[10] https://www.nytimes.com/2023/07/24/science/avi-loeb-extraterrestrial-life.html

## IV. Would any of the 2014-01-08 bolide (if interstellar) survive entry?

The 2014-01-08 meteoroid was not that big. In the spirit of recent conventions[11], it was about half the mass of a giraffe. Its mass was estimated from its measurements brightness and standard 'luminous efficiencies', which yield a kinetic energy ½ M V$^2$ = 4.6 x 10$^{11}$ J. Assuming a speed V = 44.8 km/s, this implies a mass M ≈ 450 kg. That's the mass of an iron ball only 48 cm (19 inches) in diameter.

At the terrific speeds it is assumed to have traveled at—*140 times the speed of sound*—the meteoroid would have completely vaporized in the atmosphere. The friction between a meteoroid and the air first melts its surface, causing millimeter-sized melt droplets ('ablation spherules') to spray off; but further heating can vaporize this surface, and the spherules too. At first this process is gradual, but as the meteor penetrates deeper into the atmosphere, the density of air, ρ, encountered by the meteor increases. The air pushing against the meteor creates a 'ram pressure' P$_{RAM}$ = ρ V$^2$ that eventually exceeds the strength of the material, and the meteoroid is crushed. It disintegrates into myriad fragments, each of them continuing at the same speeds and melting and vaporizing, but now with greater surface area. So much kinetic energy is suddenly converted into melting and vaporization, that the air heats and expands, creating a loud boom. The 2014-01-08 bolide suffered three such disruption events ('airbursts') in the span of 0.22 seconds. After disintegrating, the material continued traveling at 45 km/s, almost certainly vaporizing before it slowed down.

As an aside, the ram pressure at the point where the 2014-01-08 bolide disrupted was calculated by Siraj & Loeb (2022b) to be P$_{RAM}$ = 194 MPa. They then claimed this exceeded the material strength of even iron meteorites, which they said was 50 MPa, citing Petrovic (2001). It is clear Loeb *et al.* did not actually read this paper, however; even the abstract of Petrovic (2001) says "the average iron meteorite compressive strength is 430 MPa." Again, it is likely that the 2014-01-08 velocity was mismeasured, its speed was actually 22 km/s instead of 44 km/s, and that it broke up at a ram pressure ~50 MPa. That would be consistent with the strengths of more-common rocky bodies, which typically don't survive pressures greater than tens of MPa (e.g., Pohl & Britt 2020). But *if* the 2014-01-08 bolide were interstellar, and traveling at 44.8 km/s, its breakup is much more consistent with an iron meteorite.

How much of a bolide survives entry vs. vaporizes in the atmosphere depends on the material it is made of and the velocity with which it enters the atmosphere. This is determined by integrating the equations of frictional slowing and heating, and heat transfer within the meteoroid. The standard result is that the fraction of mass that does not vaporize is

$$M_F / M_I = \exp [ -\sigma (V_I^2 - V_F^2) / 2 ], \qquad (1)$$

---

[11] https://physicsworld.com/a/astronomers-launch-new-asteroid-classification-system-based-on-animal-sizes/

where $M_F$ and $V_F ≈ 0$ are the final mass and velocity, $M_I$ and $V_I$ the initial mass and velocity, and σ is the 'ablation parameter', a parameter combining information about the heat capacity and thermal conductivity, etc., of the material (e.g., Popova et al. 2019). The ablation parameter has been determined for meteors of various materials, and is typically σ > 0.01 km$^2$ s$^{-2}$ for rocky material (Revelle & Ceplecha 2001), and up to 0.07 km$^2$ s$^{-2}$ for iron (Revelle & Ceplecha, 1994).

When the ablation parameter σ is large, and especially when the initial velocity $V_I$ is large, the surviving fraction becomes exponentially small. If the 2014-01-08 bolide were interstellar and traveling at $V_I ≈ 45$ km/s—a speed significantly higher than most meteors experience—then vaporization would have been fierce. If it was a rocky body, only a fraction ≈ 4 x 10$^{-5}$, or 20 grams, could have survived entry. If it were an iron body, it would have *completely* vaporized.

In contrast, Tillinghast-Raby, Loeb & Siraj[12] (2022, arXiv 2212.00839) concluded that a significant fraction (8-21%) of the mass of the bolide could survive entry. The discrepancy could have been due to their implicitly assuming a small ablation parameter σ = Λ / (2 Q Γ). Tillinghast-Raby *et al.* assumed a heat transfer coefficient Λ = 0.04, a heat of ablation Q = 6.549 x 10$^{10}$ erg/g, and a drag coefficient Γ = 0.5, yielding σ = 0.0061 km$^2$ s$^{-2}$. This is smaller than observed for natural stony meteorites and *much* lower than for iron meteorites, so less material would be predicted to vaporize. However, we note that equations 2-6 of Tillinghast-Raby *et al*. reproduce the same equations that, when integrated, have the solution in Equation 1 above. Thus, they should have predicted that 0.2% of the mass, or 1000 g of spherules, would survive entry. We believe the authors mistakenly decreased the entry velocities of a large percentage of the fragments in their calculations.

> In summary, if the 2014-01-08 bolide were interstellar, mere grams of it would have survived entry (if rocky), and more likely all of it vaporized (if iron). Loeb *et al.* make a number of mathematical mistakes and misreadings of the literature in order to claim (in unrefereed work) that tens of kilograms might survive.

**V. Do we even know where spherules would have fallen?**

Very few spherules would have been created, but in principle these could be recovered (along with a **bunch** of background spherules), if one knew where to look. As we'll show, there are far too many uncertainties to define a reasonable search area.

Loeb *et al.* used the longitude and latitude (1.3°S, 147.6°E) and altitude (18.7 km) reported into the CNEOS database to fix the location of the fireball. In the map in their Figure 3, reproduced in **Figure 1** below, this places the fireball within a red box 0.1° (11 km) on a side. They then tried to fix the location within that box by knowing the distance *r* to a seismometer on Manus Island, about 90 km to the south. They claim this distance is *r* = 83.9 ± 0.7 km. This places the fireball within an orange band on their map. This is where they concentrated their search.

---

[12] https://arxiv.org/pdf/2212.00839.pdf

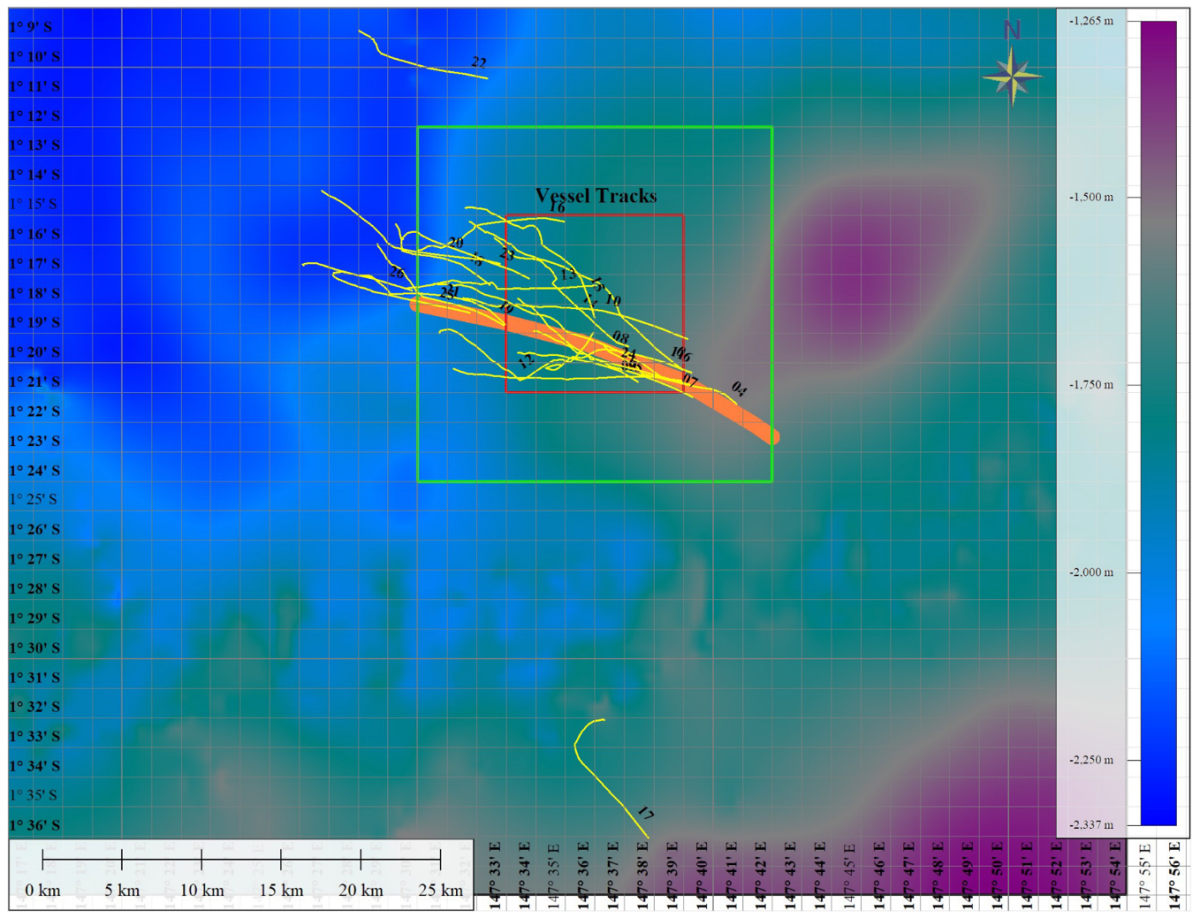

**Figure 1**. Reproduction of Figure 3 of Loeb *et al.*, showing the location where they believe the fireball occurred, according to one set of coordinates on the CNEOS website (red box) and their (flawed) estimate of the distance to the Manus Island seismometer. The actual location is much more uncertain, as is the location along the path where the spherules were produced.

There are several obstacles to this approach. One, it's not clear how accurate the reported latitude and longitude are (i.e., where to place the red box). Two, the distance to the fireball (i.e., where to place the orange band) is possibly very different and also so uncertain to be meaningless (i.e., the orange band is as wide as the red box). It's also unclear which of the three fireball disruption events is being recorded by the satellites and the seismometer, or even if they are recording the same events (where along the orange band spherules were produced).

*Location of the fireball / where to place the red box?*

The latitude and longitude on the main list of fireballs on the CNEOS website lists a location (1.3°S, 147.6°E). Uncertainties are not reported, but we presume the position is measured and reported accurately, implying uncertainties in latitude and longitude of ±5.5 km, i.e., a red box 11 km on a side centered on those coordinates. However, the location of the flash *also* is reported elsewhere on the CNEOS website[13] to have taken place at a latitude of 1.2°S, and a longitude of 147.1°E, as shown in **Figure 2**. This implies a red box 11 km on a side and centered on *those* coordinates, 11 km to the north and **55 km to the east** of where Loeb *et al.* searched.

Uncertainties in altitude are not reported by CNEOS, but comparison with ground-based observations shows the average discrepancy is ±3 km (Brown et al. 2016; Devillepoix et al. 2019).

> **Bolide Detection Notification: 2014-008**
>
> At 17:05:34 UT on 08 Jan 2014, sensors aboard U. S. Government satellites recorded the flash signature of a large meteoroid entry into the atmosphere. Location of the flash was Lat 1.2S, Lon 147.1E.

**Figure 2**. This text accompanied the light curve data on the CNEOS website. The location reported here contradicts the information reported elsewhere on the site.

*Path of the fireball?*

Presuming the velocity components Vx, Vy, and Vz reported in the CNEOS database are correct, the direction of travel of the 2014-01-08 bolide can be reconstructed. Zuluaga (2019) reported an azimuth of 285.6° (i.e., moving from the northwest, to 16° south of due east, as is evident from the geometry and negative Vz value) and an elevation of 26.8°. We note that in Figure 4 of Siraj & Loeb (2023; arXiv2303.07357), they mistakenly show the direction of travel of the meteor being exactly in the opposite direction.

The direction of travel is the most uncertain quantity for fireballs because the velocities are so uncertain (Devillepoix et al. 2019). For example, if we assume Vy was mismeasured and was -19 km/s (as above), the azimuth would be off by 12°. The travel directions of several meteors in the CNEOS database are off by as much as 90° (Brown et al. 2016; Devillepoix et al. 2019). *We may have no practical idea of the meteor's path.*

As the meteor traveled along its path, it suffered three disintegration events, with the first and last spaced about 0.22 s apart, as shown by the light curve depicted in **Figure 3**. If traveling at 45 km/s, the first and third events would have been separated by 9 km on the ground, and 4.5 km in altitude. It is not clear whether the position reported in the CNEOS database refers to the first disruption event at 0.35 s, or the last and brightest event at 0.57 s. If it's the first, but most spherules and the earliest-arriving sound waves are generated by the third, then the seismometer data is referring to an event 9 km to the east of the location reported in the CNEOS database (whichever set of coordinates is right).

---

[13] https://cneos.jpl.nasa.gov/fireballs/lc/bolide.2014.008.170534.pdf

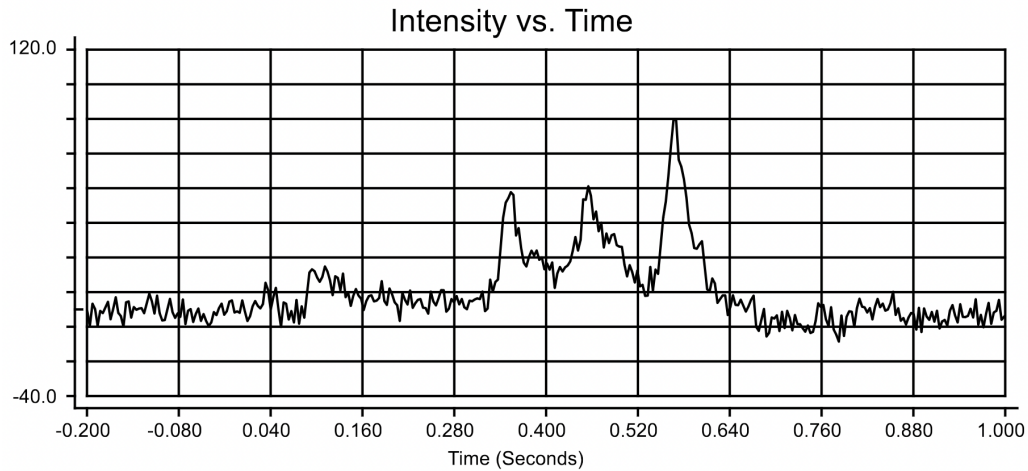

**Figure 3**. Light curve of the 2014-01-08 bolide reported by CNEOS.

*Distance between the Manus Island seismometer and "the" fireball?*

Siraj & Loeb (2023; arXiv2023.07357)[14] attempted to fix the distance to the fireball (more precisely, to one of the three disintegration events) using the timing of the first sound wave from the fireball to reach a seismometer on Manus Island in Papua New Guinea (located at 2.0432°S, 147.3662°E). This seismometer is roughly 86 ± 6 km away, if the fireball were at 1.3°S and 147.6°E; or 97 ± 6 km away, if the fireball were at 1.2°S and 147.1°E. If the sound speed were 347 m/s, the signal would arrive about 248 ± 17 s or 280 ± 17 s later. This gives a ~1 minute window in which sound waves might arrive at the seismometer. If the seismometer recorded a signal in this window, it might be attributed to the fireball, and the time of arrival of sound waves used to refine the distance between the seismometer and "the" fireball.

As an aside, Loeb *et al.* state these locations are outside the 22 km limit for the "territorial waters" of Papua New Guinea; but they clearly are within its Exclusive Economic Zone, as set by the United Nations Convention on the Law of the Sea (UNCLOS). Papua New Guinea officials have accused the authors of removing materials without permits, as described in *The Post* on July 10, 2023[15].

Refining the location of the fireball necessitates confidently identifying a shaking of the seismometer with the arrival of sound waves from the fireball. This cannot be done with fully certainty. **Figure 4** shows the signal from the Manus Island seismometer, downloaded from the International Federation of Digital Seismograph Networks website[16]. The event attributed to the fireball is labeled. Evidently, events comparable in seismic energy to the purported fireball event occur on average every 5-10 minutes or so, including a similar event that occurred only 4 minutes earlier. Within the ~1 minute window in which Siraj & Loeb sought a signal from a sound wave, there is a decent chance (10-20%) that a background event would occur.

---

[14] https://arxiv.org/pdf/2303.07357.pdf
[15] https://www.thepost.co.nz/a/world-news/350033283/scientists-snatched-our-alien-beads-say-authorities-in-papua-new-guinea
[16] https://www.fdsn.org/networks/detail/AU/

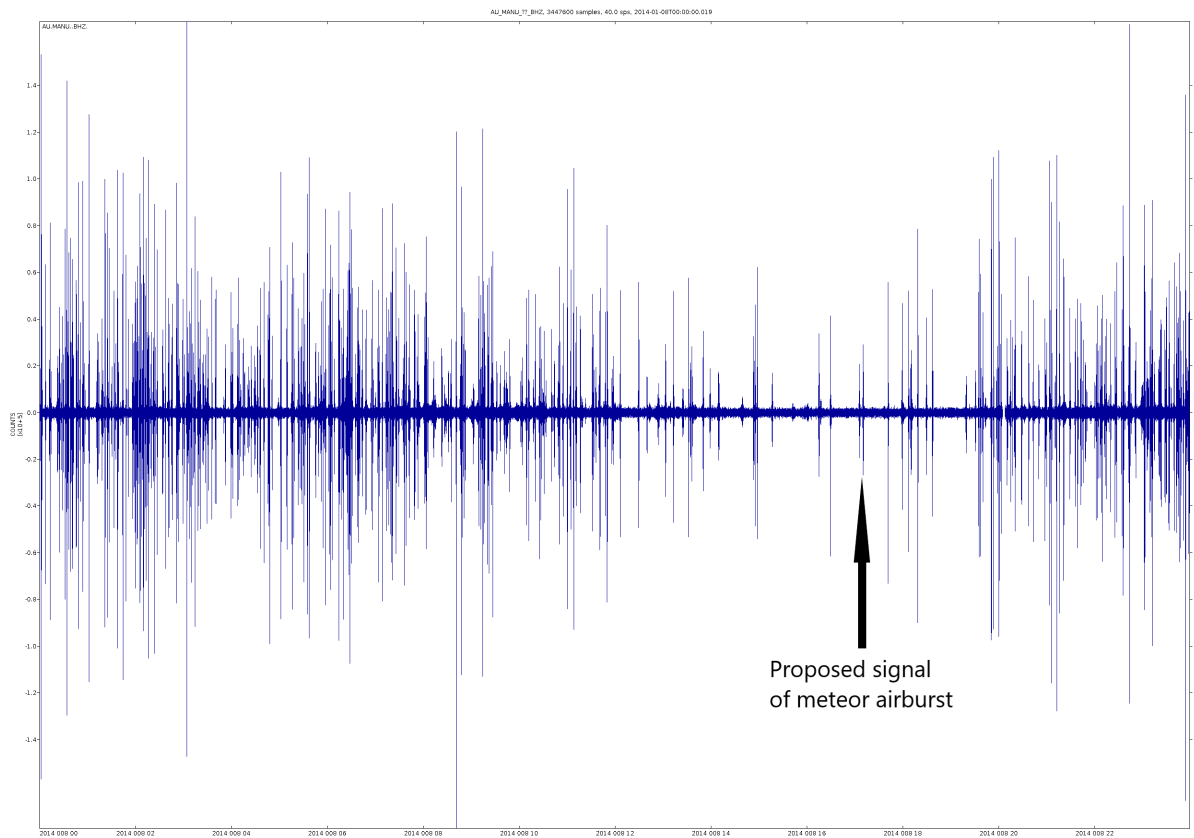

**Figure 4**. Data from the Manus Island seismometer during the 24-hour period of Jan 8, 2014.

Siraj & Loeb identified the labeled event with the fireball and essentially assumed that the distance to the fireball is an average sound speed $V_A$ (about 347 m/s near the ground) multiplied by the time after the flash for the sound waves to reach Manus Island, $\Delta t \approx 270$ s. They concluded that the ground-distance between the seismometer and the fireball (at least, the nearest or loudest one) was $r$ = 83.9 ± 0.7 km. Based on that presumed uncertainty, the orange band in Figure 3 of Loeb *et al.* is made only 1.4 km thick, much smaller than the red box. However, this uncertainty, only ±0.8%, is remarkably small considering all the complications. They also concluded the altitude of the fireball was 16.9 ± 0.9 km. It is curious that they accept this discrepancy between this value and the 18.7 km reported in the CNEOS database.

Siraj & Loeb provide hardly any details about how the arrival times of sound waves at the seismometer were calculated, so it is difficult to judge its correctness. One issue is that, because the sound speed varies with temperature and therefore height, as $V_A$ = [ 347.2 − 1.861 ($z$ / km) ] m/s, sound waves are refracted away from the ground or ducted through the stratosphere. Notably, Siraj & Loeb did not recognize that this formula does not apply in the stratosphere (z > 10 km), where the sound speed is uniform with height; this makes a difference of several seconds in arrival times, and at least 1 km in distance.

Because we could not check the details of their model, we reproduced their calculation as well as they described it. We find that the first sound waves would arrive at the seismometer 270.0 s after the fireball if they traveled a distance $r$ = 88.5 km along the ground after originating at an altitude of 18.7 km; or a distance $r$ = 89.2 km if they originated at 16.9 km. These values of $r$ differ from those of Siraj & Loeb; we would shift the position of the orange band 5 km to the north of where Loeb *et al.* searched. Either we and Siraj & Loeb are making different assumptions, or one or both of us has miscalculated. Either way, reasonable differences about how the calculation is conducted probably lead to differences in $r$ of about 5 km.

This calculation also shows there is a trade-off between ground-distance and altitude; for every 1 km variation in the fireball's altitude, there is a 0.4 km difference in ground-distance. Altitudes of fireballs in the CNEOS database are uncertain at the ±3 km level, and the first and last disruption events differed in altitude by 4.5 km. This translates into additional uncertainties in $r$ of ±1 km, and possibly a shift in 2 km in where to place the path on the ground.

The biggest source of uncertainty, though, concerns the arrival time of sound waves after the flash. Siraj & Loeb assumed Δt = 270.0 ± 0.5 s, a very small uncertainty. Reported times of fireballs in the CNEOS database are accurate to "within a few seconds" according to Devillepoix et al. (2019). In addition, inspection of **Figure 5**, which shows the intensity of sound waves at the seismometer, suggests the uncertainty is at least ±5 seconds. This translates into additional uncertainties ±2 km of where to place the path.

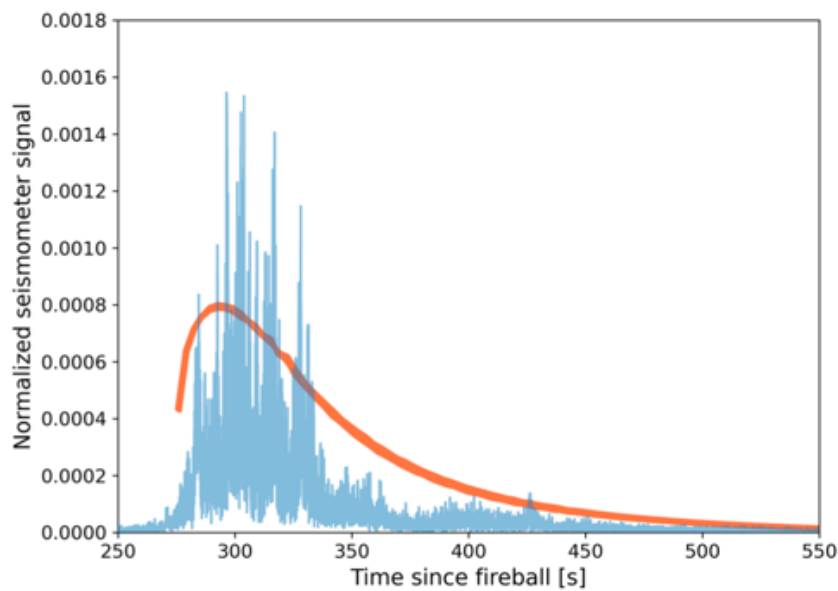

**Figure 5**. Figure 3 from Siraj & Loeb (2023). The blue curves denote the intensity of sound waves at the Manus Island seismometer. The orange curve is an unexplained model. Did the first signals arrive at 260 s, 270 s, or 280 s?

An additional issue is, what would sound waves provide a distance to? As depicted in Figure 3, there were *three* disruption events that could produce explosions and sound waves. The path of the bolide had an azimuth of 16° south of east (assuming the velocities are correct), while Manus Island lay roughly 17° west of south, so that the bolide traveled almost perpendicular to the line of sight to the seismometer. In this geometry, the first sound waves would arrive at the seismometer from the third and final disruption event, because it is at the lowest altitude and also the loudest. However, if the velocities are incorrect and the path substantially to the northeast instead, then sound waves from the first disruption event would arrive first.

There is an additional factor to consider: solid material generated atmospheric entry tends to be dropped along a long path called a strewn field. Strewn fields can be very long (> 10 km), and that certainly would be the case here, if each of the three disintegration events dropped spherules along the path. Strewn fields are also very wide: > 2 km for iron meteorites, and > 4 km for stony meteorites (e.g., Moilanen et al. 2021). In addition, spherules would take varying times ($\approx$ 1-30 minutes) to sink to the seafloor, during which time the Equatorial UnderCurrent would sweep them westward several km (Desch et al. 2023).

In summary, the distance is very uncertain, and the thickness of the orange band where spherules could be found should be about 4 times wider than Loeb *et al.* drew, almost as large as the red box. In addition, the orange band may need to be shifted 5 km to the north. If the flash reported in the CNEOS database refers to the first disruption event, then the meteor would have traveled roughly 9 km to the east before generating most of the spherules. All in all, the uncertainties are basically comparable to the size of the red box.

And of course, there is still the issue that red box may be *55 km to the west*, depending on which coordinates from the CNEOS website one uses. The red box would be at a distance of 97±6 km, and my estimate for the distance of the orange band, 89±3 km, would allow some overlap ($r \approx$ 92 km) at that location, so that is also a viable solution to all the available data.

*Summary*

The uncertainties in when sound waves would arrive at the Manus Island seismometer make the distance *r* to the fireball almost as uncertain as the error in the reported position, ±5.5 km. The trajectory of the fireball, being perpendicular to the line-of-sight to Manus Island, makes it difficult to fix where along the path the bolide disintegrated; the search area could be shifted 9 km to the east from the fireball. Of course, two locations are reported in the CNEOS database, 55 km apart, and both are consistent within uncertainties with the seismometer data. It's just incredibly naïve and overconfident to claim to know where spherules would have fallen with sufficient accuracy to mount a search.

**VI. Did the concentration of spherules vary across the expedition site?**

Loeb *et al.* claimed that the density of spherules (number per km$^2$) was much higher in the regions they associated with the path of the 2014-01-08 bolide than in regions definitely off the path. They argue that a significant fraction of the spherules they recovered along the assumed path therefore must have come from the bolide. They specifically claim: "Given that the highest-yield (yellow) regions in the heat map account for roughly twice the background yield owing to the excess spherules added by IM1, we expect to find a significant numbers [sic] of IM1's spherules relative to background spherules in the highest-yield regions." However, Loeb *et al.* collected far too few spherules to make this claim. It also would be wildly unexpected for even a small fraction of spherules collected to be from the 2014-01-08 bolide.

*Density of background spherules*

The ocean floor is littered with a background concentration of cosmic spherules and micrometeorites, as well as volcanic ash and anthropogenic coal ash. Cosmic spherules have been collected from the seafloor for decades (e.g., Millard & Finkelman 1970; Blanchard et al. 1980; Brownlee 1985; Rudraswami et al. 2011; Prasad et al. 2015; etc.) Millard & Finkelman (1970) collected 1200 spherules from 750 kg of Pacific red clay. Assuming both they and Loeb *et al.* collected material from the same depths (~10 cm), that equates to roughly 4 x 10$^6$ spherules per km$^2$. The flux of micrometeorites is uniform across the globe, so any seafloor site with similarly slow deposition rates ~1 mm/kyr (including this one) should have similar and uniform background concentrations.

As an aside, Loeb *et al.* collected from 0.26 km$^2$ area, meaning they must have been ~10$^6$ spherules in their search area. The fact that they gathered only ~700 suggests a surprisingly low collection efficiency, perhaps because their collection was skewed to the very largest particles.

*Spatial density of spherules from the 2014-01-08 bolide*

The additional contributions from the 2014-01-08 bolide would have been small in comparison to the background. Assuming a typical spherule mass ~1 mg, at most ~1 x 10$^6$ spherules would have been created, if 0.2% of the bolide survived vaporization. These would have been spread out over a path at least > 10 km long and > 4 km wide (the typical dimensions of strewn fields), not even accounting for the additional spreading due to ocean currents. This yields a maximum density < 2 x 10$^4$ spherules per km$^2$. Even assuming the most generous ablation parameter, so that 0.2% of the mass survives vaporization, spherules from the bolide would make up < 1% of the background. But for a realistic ablation parameter, far, far less of the bolide would survive.

There is every expectation that > 99% (probably 100%) of the spherules Loeb *et al.* collected were simply part of the background of micrometeorites deposited on the seafloor over the last tens of thousands of years.

*Are there more spherules in the presumed path of the bolide?*

Loeb *et al*. have argued that the concentration varied spatially, being about twice as high in the regions they associate with the path of the 2014-01-08 bolide than in the two "control" regions corresponding to tracks 17 and 22 in their Figure 3. The path of the bolide, we've argued, is more uncertain that Loeb *et al.* realize; but it is fair to say that Tracks 17 and 22 probably do not include spherules from the 2014 bolide, and reflect the background concentration. Are the concentrations in tracks 17 and 22 lower than in the other regions?

Loeb *et al.* report that 622 spherules were collected in runs 1-24: 105 in runs 17 and 22, and 517 in the other runs. Unfortunately, Loeb *et al.* did not provide any useful information about the collection area per run, including for how long they dragged the seafloor, or where along each run the spherules were collected. Instead, one must estimate the collected spherules per area of seafloor from the map in their Figure 3. Roughly 11 pixels of area (each pixel in Figure 3 has area 3.43 km$^2$) can be associated with runs 17 and 22, meaning there are 2.8 collected spherules per km$^2$. About 56 pixels of are associated with the other runs, yielding or 2.7 collected spherules per km$^2$. Even in regions sampled by runs 13, 14, and 15, supposedly of high spherule concentration (as shown by the heatmap density depicted in their Figure 5), there are 183 spherules in 20 pixels of area, or 2.7 spherules per km$^2$. Within the uncertainties (±10% from counting statistics alone), these are the same. It's not clear how the heatmap was calculated, what data went into it, or what it's measuring, but there is no evidence of spatial variations in spherule concentration in or out of the assumed path of the 2014-01-08 bolide.

Even if there had been variations, it's not clear they could be attributed to the bolide rather than variations in sampling depths, collection efficiencies, deposition times, etc.

> In summary, Loeb et al. did not hit some sort of motherlode of spherules from the 2014-01-08 bolide. Even under the most generous assumptions about survival of material, it would have barely made a dent in the abundances of spherules. Almost all, if not literally all, of the cosmic spherules collected by Loeb *et al.* are from the background population of micrometeorites deposited on the seafloor over the last tens of thousands of years.

**VII. What do the Fe isotopes mean?**

Loeb *et al.* measured the iron isotopic ratios $^{57}$Fe/$^{54}$Fe and $^{56}$Fe/$^{54}$Fe in 9 of the spherules (and in S21 twice). They found them to array along a line of slope ≈3/2 (1.46) in a "three-isotope" plot. They interpret spherules with high $^{57}$Fe/$^{54}$Fe and $^{56}$Fe/$^{54}$Fe to have lost Fe by vaporization during atmospheric entry, and conclude they formed in a fireball. They miss that the slope itself is not diagnostic of vaporization *per se* and that the magnitudes of the deviations from terrestrial iron isotopic ratios suggest loss *or gain* of Fe by chemical means other than vaporization. Most importantly, the line connecting the spherules passes directly through the terrestrial values, indicating that almost all the iron in these spherules is **from our Solar System**.

**Figure 6** reproduces Figure 12a of Loeb *et al.* and shows the fractional deviations of Fe isotopic ratios from a terrestrial standard. Here, $\delta^{56/54}Fe = 1000 \times [\ (^{56}Fe/^{54}Fe)_{sample} - (^{56}Fe/^{54}Fe)_{standard}\ ] / (^{56}Fe/^{54}Fe)_{standard}$, and $\delta^{57/54}Fe = 1000 \times [\ (^{57}Fe/^{54}Fe)_{sample} - (^{57}Fe/^{54}Fe)_{standard}\ ] / (^{57}Fe/^{54}Fe)_{standard}$. Samples that match the terrestrial standard exactly have $\delta^{56/54}Fe = 0$ and $\delta^{57/54}Fe = 0$. These quantities can be used to diagnose chemical processes. If a sample loses Fe by any chemical reaction, including vaporization, will tend to lose its lightest isotopes fastest, as chemical reaction rates vary as $1/(m_{Fe})^{1/2}$. As such, loss of Fe will make the $^{56}Fe/^{54}Fe$ ratio increase, but it will make the $^{57}Fe/^{54}Fe$ ratio increase even more. In a plot of $\delta^{57/54}Fe$ vs. $\delta^{56/54}Fe$, samples that started with a terrestrial value (at $\delta^{56/54}Fe = 0$ and $\delta^{57/54}Fe = 0$) will move along a line of slope ≈1.48, called the 'terrestrial fractionation line' or 'TFL', to the upper right. Samples that gained Fe through a chemical process will move along the line to the lower left. The deviations typically are in parts per thousand, 'permil', symbolized ‰.

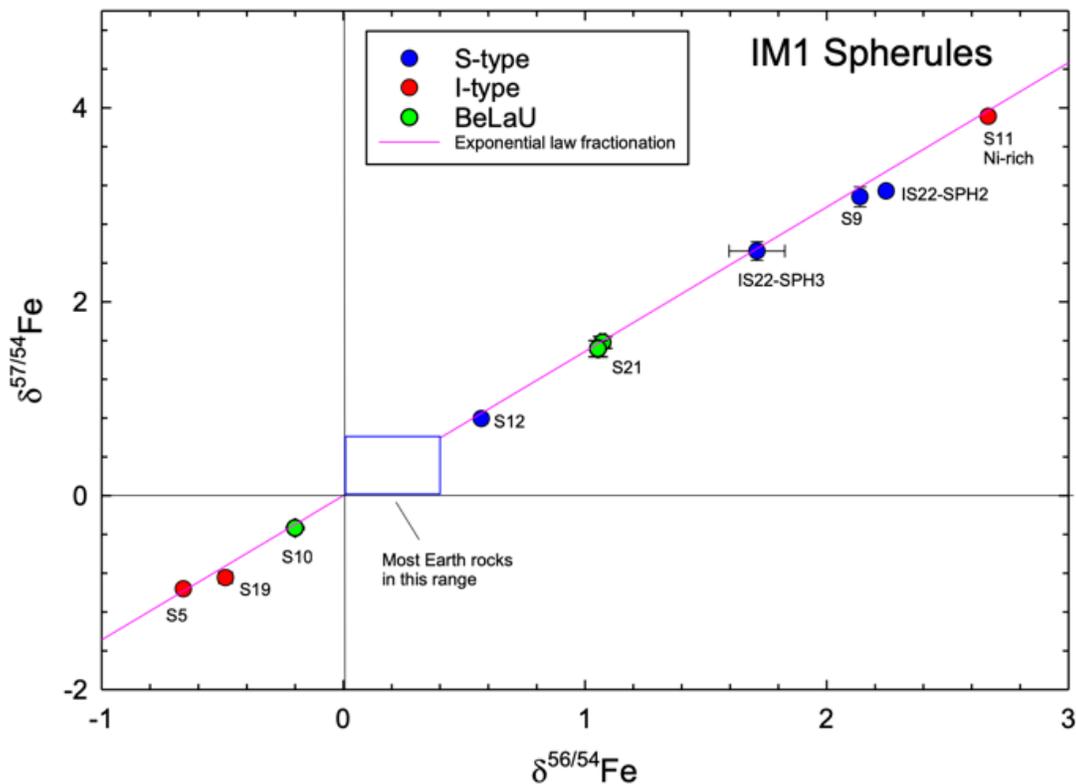

**Figure 6.** Deviations of iron isotopic ratios from terrestrial ratios in nine spherules, reproduced from Figure 12a of Loeb *et al.* The pink line adjoining them is the terrestrial fractionation line (TFL) on which samples should array if they have a Solar System origin and experienced chemically-driven mass-dependent isotopic fractionation by gain or loss of Fe, for example by vaporization of Fe. **All nine spherules are exactly consistent with a Solar System origin**.

Figure 6 shows the $\delta^{56/54}Fe$ and $\delta^{57/54}Fe$ values for nine spherules, including two of the five "BeLaU" spherules. All are found to lie on the TFL (the pink line), on which all samples on Earth lie. Indeed, within the blue box lie most samples from Earth, Mars, Vesta, and other meteorites; all samples from our Solar System lie on the TFL. This raises two questions. First, why do these samples, purportedly from another solar system, have Fe isotopes that are such an exact match to Solar System materials? Second, why is it that, even compared to Solar System materials, they have such a small range of $\delta^{56/54}$Fe values between -1 and +3?

*Why are the spherules' $\delta^{56/54}$Fe and $\delta^{57/54}$Fe so close to Solar System values?*

Figure 6 is quite damning against the interstellar hypothesis. If the "BeLaU" spherules truly were samples of another stellar system, they would not have Fe isotopic ratios so close to terrestrial. While all Solar System materials lie very close (<< 1‰) to the TFL, one would expect materials in another system to have different starting values (hundreds of ‰ away from the TFL), reflecting their time and place of formation in the Galaxy.

Because there is no reason to expect a meter-sized iron meteoroid to be destroyed over billions of years, an interstellar meteoroid could originate in system from almost any time over the Galaxy's 12 Gyr history, from any star orbiting at about the same distance (within ~1 kiloparsec) as the Sun around the Galaxy. Galactic Chemical Evolution (GCE) models show that materials forming at various times and places would sample metallicities (heavy element-to-hydrogen ratios) $Z$ and combinations of $\delta^{57/54}$Fe and $\delta^{56/54}$Fe very different from the Sun. Stars in the Milky Way over the last 7 Gyr sample metallicities $Z \approx 0.5 - 1.3 \times Z_{Sun}$ (Carrillo et al. 2023). According to Kobayashi et al. (2011), such stars should sample $\delta^{56/54}$Fe ≈ -100 to +1500, and $\delta^{57/54}$Fe ≈ 0.9 $\delta^{56/54}$Fe ≈ -100 to +1300. Stars with any Fe isotopic ratios across these ranges could contribute to the background of interstellar meteoroids.

Assuming any star formed over the last 7 Gyr could contribute equally to interstellar meteoroids, there would be a < 5% chance its $\delta^{56/54}$Fe value would lie within 50‰ of the Sun's value (a generous maximum variation due to vaporization or aqueous alteration). More importantly, there would be a < 0.1% chance, whatever its $\delta^{56/54}$Fe value, that its $\delta^{57/54}$Fe value would lie within < 1‰ of the TFL. On the face of it, then, < 0.005% of all systems formed over the last 7 Gyr would produce materials whose Fe isotopes so closely matched Solar System values. A more careful analysis will show the exact fraction to be different from 0.005%, but it will surely be similarly small.

It thus would appear there is a > 99.995% chance that the spherules measured here, including the "BeLaU" spherules S21 and S10, reflect a Solar System Fe isotopic ratio. There is only a < 0.005% chance another solar system would match ours so exactly.

*Variation of spherules' $\delta^{56/54}$Fe*

Loeb *et al.* argued that spherules in Figure 6 array along the TFL because of loss of Fe during vaporization. Rayleigh distillation (loss of Fe by vaporization) has the effect of increasing $\delta^{56/54}$Fe and moving samples up along the TFL. Cosmic spherules of Solar System origin mostly derive from materials at $\delta^{56/54}$Fe ≈0 and $\delta^{56/54}$Fe ≈0 (placing them on the TFL); after vaporization, S-type (rocky) spherules tend to have $\delta^{56/54}Fe \approx +1.4$ to $+3.2$, and I-type (iron) spherules tend to have $\delta^{56/54}Fe \approx +20$ to $+36$.

Isotopic shifts due *only* to vaporization can be ruled out. Notably, the "BeLaU" spherule S10 and the I-type spherules S5 and S19 have $\delta^{56/54}Fe < 0$. All Solar System sources (including Mars, angrites, Vesta, ureilites, etc.) happen to lie on the TFL, but because none have $\delta^{56/54}Fe < 0$, the three spherules S10, S5 and S19 could not derive from Solar System materials (if one only considers vaporization). These three spherules would have to derive from the same extrasolar body, presumably 2014-01-08. It is exceedingly improbable that this source would like on the TFL so close to terrestrial values (see above), and it's also very unlikely that all these samples would lie on the same line if only half were from the bolide (as Loeb *et al.* suggest); but also the magnitude of the variations in $\delta^{56/54}Fe$ are too small to be consistent with vaporization alone.

The $\delta^{56/54}Fe$ values of the S-type, I-type, and "BeLaU" spherules all lie between -1 and +3, a spread of only 4‰. In contrast, the spread among cosmic spherules is much larger: 16‰ among the I-types alone, and about 35‰ between the S-type and I-type spherules. These variations have to do with how Fe vaporizes from these samples, and even interstellar materials should see similar ranges of $\delta^{56/54}Fe$ values. This strongly suggests that all of these spherules later exchanged about 90% of their Fe atoms with a terrestrial reservoir near $\delta^{56/54} \approx 0$, after falling into the ocean. Presumably they reacted with seawater.

Reaction with seawater also seems consistent with the correlation of $\delta^{56/54}Fe$ values with Fe/Ni ratios, suggesting they lost or even gained Fe, *after* atmospheric entry. In terrestrial samples, samples with $\delta^{56/54}Fe < 0$ tend to result from *gain* of Fe by aqueous alteration, as in terrestrial magnetite with negative $\delta^{56/54}Fe$. Gain and loss of Fe is suggested among the I-type spherules analyzed here. S11, with positive $\delta^{56/54}Fe$, has lower Fe abundance (3.25x CI) and a lower Fe/Ni ratio (0.62x CI). That's consistent with some mild loss of Fe during atmospheric entry. Spherules S5 and S19, with negative $\delta^{56/54}Fe$, have higher Fe abundances (3.75x CI) and have very high Fe/Ni ratios (145x and 1880x CI). Such extreme fractionation of Fe and Ni is not expected during vaporization. All of this strongly suggests gain or loss of Fe by aqueous alteration instead.

> In summary, the Fe isotopic ratios of the spherules definitively identify them as from our Solar System, and not an extrasolar source. They could have experienced isotopic fractionation during vaporization, moving them up the TFL, but this cannot be the only effect. Their Fe isotopic ratios appear most consistent with massive exchange of Fe with a terrestrial water, presumably via aqueous alteration while sitting on the seafloor for tens of thousands of years.

**VIII. What does it mean that S21 is a triple compound spherule?**

Some of the spherules, including one of the "BeLaU" spherules (S21), are found to be compound, the fusion of three smaller spherules while still partially molten. Loeb *et al.* argue "the existence of a triple-merger like S21 can be explained as a product of IM1's airburst." and they provide a back-of-the-envelope calculation to support this claim. However, their calculation is flawed, and if it were true, then compound spherules like S21 would be common, which they are not. The existence of compound spherules like S21 is actually proof that the "BeLaU" spherules did **not** form in an airburst like the 2014-01-08 bolide.

It is possible, in theory, for some ablation spherules coming off the bolide to collide while in mid-air before cooling off. Loeb *et al.* estimate the probability of two spherules colliding to be $\tau \sim n \sigma R \sim 0.3$, assuming the number density of spherules is $n \sim 0.2$ cm$^{-3}$, the cross-sectional area is $\sigma \sim 0.03$ cm$^2$, and the typical distance traveled through a cloud of spherules is $R \sim 50$ cm. The probability of two collisions, producing a triple spherule, would then be $\tau^2 \sim 0.1$, suggesting spherules would be common in the 2014-01-08 bolide and all others.

In fact, compound spherules are very rare among cosmic spherules, and it is notable when they are found, such as in the impact event in Antarctica 430 kyr ago (van Ginneken et al. 2021). Their presence in that event appears to require formation in a very large impact plume, not a typical fireball. This implies S21 was not formed in the 2014-01-08 fireball, either.

Indeed, the back-of-the-envelope calculation of Loeb *et al.* seems fundamentally wrong. First, the spherules have an average diameter 600 µm, mutual cross sections $\sigma \sim 0.01$ cm$^2$, and masses about 1 mg. The entire mass of a 500 kg meteoroid could be converted into $5 \times 10^5$ such spherules. But a number density as high as $n \sim 0.2$ cm$^{-3}$ would require they be distributed in a volume no larger than 2.5 m$^3$, a sphere with radius ~1 m. While this is consistent with the assumption by Loeb *et al.* that spherules travel ~50 cm through a cloud of other spherules, it is certainly wrong.

Spherules would have been generated during the 0.22 s the 2014-01-08 bolide was ablating, during which they would have escaped into a volume with cross-sectional area 2000 cm$^2$ (the meteoroid's cross-sectional area) times the length of the channel, 10 km (how far the meteoroid traveled in 0.22 s). The channel has volume 2000 m$^3$, and the number density *n* of spherules therefore is three orders of magnitude smaller than Loeb *et al.* assumed. A value of $R \sim 1$ m is appropriate, but only because spherules have at most $t \sim 1$ s to collide with another spherule before they cool and solidify, and the proper value of $R$ is $V_{REL} t$, where $V_{REL} \sim 1$ m/s is the maximum relative velocities between spherules that allows them to collide and stick instead of shattering. The net result is that the probability of collision is *at most* $\tau \sim 3 \times 10^{-6}$, even if *none* of the material vaporized. The probability of a triple spherule would be a vanishingly small $< 10^{-11}$.

> In summary, a proper calculation shows why compound spherules are so rarely produced in regular fireballs, and practically rules out a fireball origin for the "BeLaU" spherule S21. It is likely to have an origin in a relatively rare impact plume, like the one in Antarctica, which occur on ~$10^5$ yr timescales (von Ginneken et al. 2021). Very likely, S21—and by extension, other "BeLaU" spherules—have sat on the ocean floor for tens of thousands of years.

## IX. Are the "BeLaU" compositions new and unique?

Loeb *et al.* analyzed the chemical compositions of 57 spherules and found that many did not match their expectations, including a subset of 5 spherules enriched in Be, La and U that they call "BeLaU" spherules. Loeb *et al.* assert "these have never been described in the cosmic spherule literature." Their argument that these have an extrasolar origin hinges entirely on there not being **any** examples of similar compositions among cosmic spherules from asteroidal micrometeorites in our Solar System. It is not clear how exhaustive their search of the entire cosmic spherule literature was; it appears to have been limited to the single paper by Folco and Cordier (2015). It was shockingly easy to find examples in the literature of (Solar System) spherules with surprisingly similar compositions, indicating that the "BeLaU" spherules almost certainly have a Solar System origin as well.

Anthropogenic coal ash is a very likely candidate, as recently pointed out by Gallardo (2023). Even half a century ago, 5% of all magnetic spherules on the mid-ocean seafloor were coal ash, and 60% of magnetic spherules near land (Parkin et al. 1970); these fractions are surely higher now. It has been recognized for just as long that diagnosing between coal ash and cosmic spherules is difficult and cannot be done morphologically, and requires a comprehensive chemical analysis (Doyle et al. 1976). It is stunningly negligent that Loeb *et al.* did not present a plan for distinguishing between cosmic spherules and coal ash, or even consider this possibility.

Loeb *et al.* analyzed the chemical compositions of 57 spherules. They found that 18 matched the "S-type" (silicate-rich) spherules and 18 more matched the "I-type" (iron-rich) spherules described by "Folco et al. (2015)" (they meant Folco and Cordier 2015). Another 21 did not match types specifically described by "Folco et al. (2015)", and—it bears repeating—at that point the authors concluded "these have never been described in the cosmic spherule literature." They then proceeded to assert that the chemical compositions of these spherules, being depleted in Mg, "are clearly derived from material that has gone through planetary differentiation." Ignoring the ability of sea water to leach Mg out of minerals, Loeb *et al.* give these a new name ("D-type"). Some are depleted in La (relative to CI chondrites) and others (including the spherule S21) are enriched in Be, La, and U; the latter they call "BeLaU".

What is this unusual chemical pattern? **Figure 7** shows Figure 10a of Loeb *et al.*, with the mass fractions of various elements in the five "BeLaU" spherules, normalized to their abundance in CI chondrites. CI chondrites are the most primitive types of meteorites, matching the composition of the Sun. Other chondrites (meteorites from unmelted asteroids) roughly match CI chondrites in composition. Any number of processes can alter compositions of meteoritic materials, including vaporization, melting, aqueous alteration, etc. The elements in Figure 7 are arrayed from left to right in order of increasing volatility (ability to be vaporized).

It is common in the meteoritics literature to report molar *ratios* of elements relative to a chemically similar element, normalized to the same ratio in CI chondrites, e.g., $(U/Lu)_{sample} / (U/Lu)_{CI}$. (Uranium and lutetium are similarly non-volatile, rock-loving elements.) In contrast, it is common in the cosmic spherules literature to report mass fractions of elements relative to the mass fractions in CI chondrites, i.e., $X(U)_{sample} / X(U)_{CI}$. The difference matters because mass fractions are easier to change than elemental ratios. For example, 40% of the mass of CI chondrites are Mg, Si, and Ca, and the O atoms that go with them (Lodders 2003). These elements are also easily leached out of minerals by water. Losing most of the Mg, SI and Ca from a CI chondrite would raise the mass fractions of U, Lu, etc., by a factor of 1.6, even if they didn't participate in any chemistry. The (U/Lu) ratio, in contrast, would be unchanged.

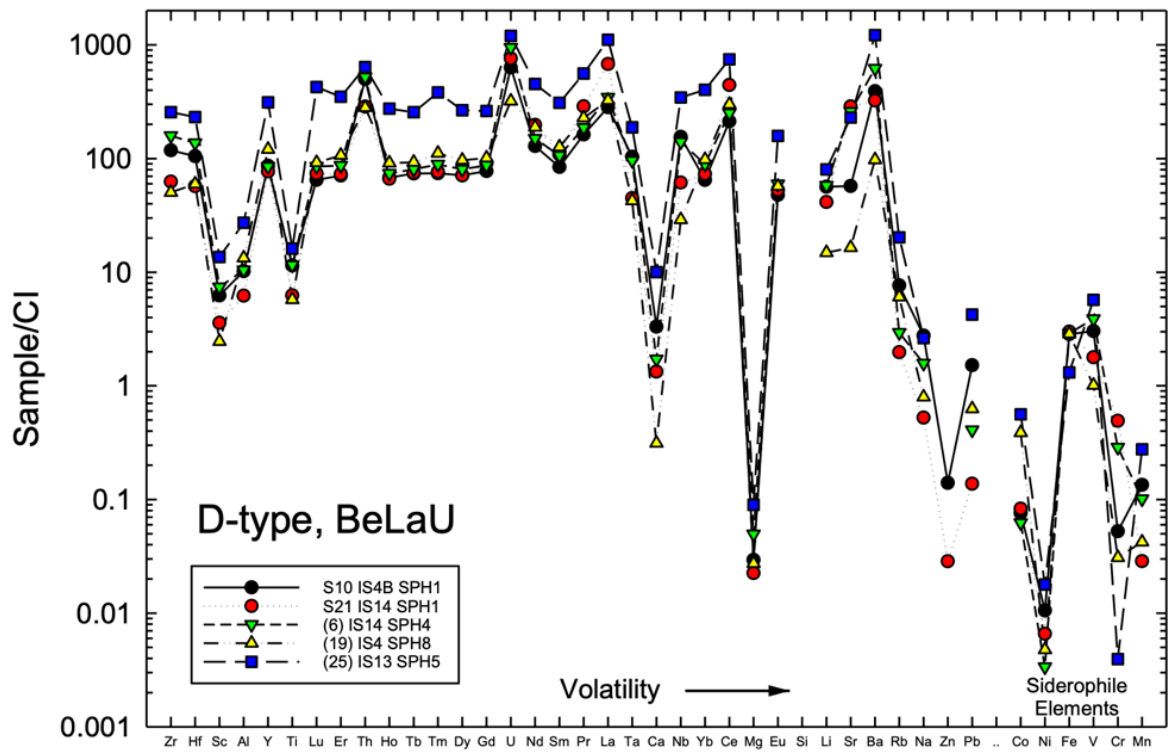

**Figure 7**. A reproduction of Figure 10a of Loeb et al., showing the mass fractions of various elements in the five "BeLaU" spherules, normalized to the mass fractions of the same elements in CI chondrites.

With this in mind, we focus first on the elemental abundance *pattern*, the ratios of elements with respect to each other. Most rare earth elements (REEs) like Lu, Er, etc., are present at 80× CI in spherule S10, and 300× CI in spherule S21, but in both, the relative enrichments are: U/REE ≈ 7.5× CI; La/REE, Ba/REE, Th/REE all ≈ 4× CI; plus slight enrichments in Ce, Hf, and Pb. The pattern also includes severe depletions in Mg, with MgO mass fractions ≈ 0.5wt%.

Delving into the causes of these patterns later, the immediate question is whether similar abundance patterns exist in other cosmic spherules. It turns out that a very similar pattern, reproduced in **Figure 8,** is found in spherules recovered from the Indian Ocean by Rudraswami et al. (2016). For example, in the relict-bearing spherules, most REEs are present at levels ≈ 2× CI. The most enriched elements are (in order): Th (Th/REE ≈ 10× CI), U (U/REE ≈ 6× CI), Ba (Ba/REE ≈ 6× CI), La (La/REE ≈ 4× CI), then Ce, Hf, and Pb. The MgO mass fraction is ≈35wt%.

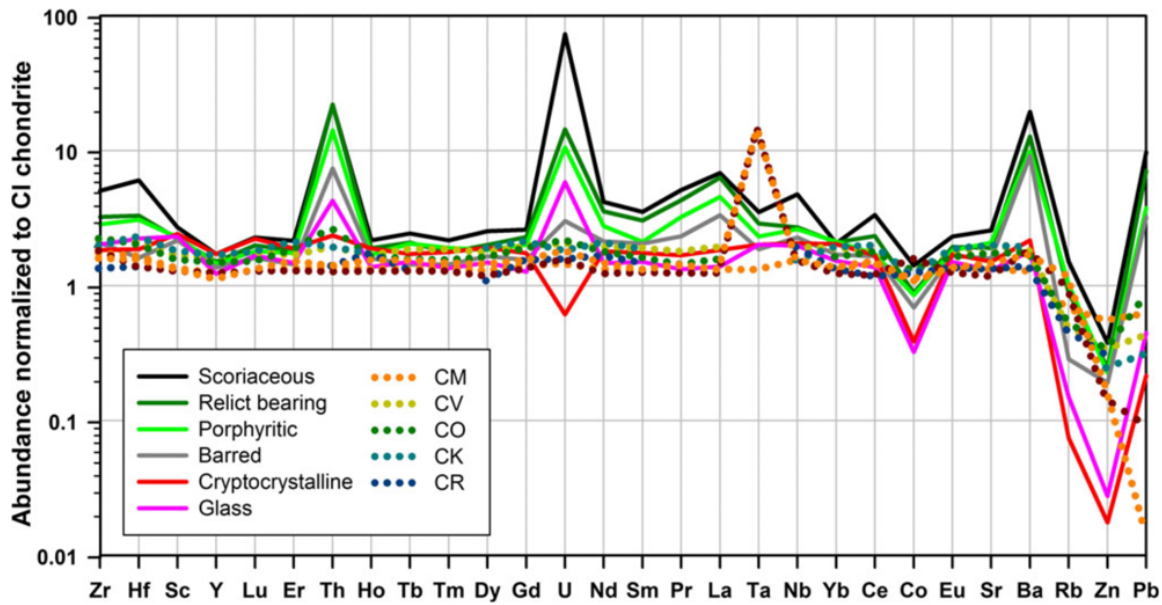

**Figure 8**. Mass fractions of elements in cosmic spherules recovered from the Indian Ocean, normalized to mass fractions in CI chondrites, from Figure 2 of Rudraswami et al. (2016).

These patterns match remarkably well. They have the same maximum enrichments in the same four elements (U, La, Ba, Th). The enrichments are similar, on the order of 4-10× CI. The elements enriched to lesser degrees are the same: Ce, Hf, Pb.

As for the absolute mass fractions, the Indian Ocean spherules have 35wt% MgO, compared to 0.03wt% in the "BeLaU" spherules. The Papua New Guinea spherules appear to have been depleted in Mg and other major rock-forming elements, by a factor of 100 relative to the Indian Ocean spherules. It follows that the mass fractions of REEs, La, U, Th, Ba, etc., would be elevated by a factor of 100 times.

Notably, micrometeorites from the impact plume in Antarctica 430 kyr ago also are enriched in La, Th, and U, attributed to terrestrial contamination (van Ginneken et al. 2021).

> In summary, the claim that these abundance patterns "have never been described in the cosmic spherule literature" is laughable. Loeb *et al.* failed to do due diligence in their literature review. This is a chemical pattern seen in spherules in very different parts of the world and are not unique to the spherules from the 2014-01-08 bolide (if they could even be associated with it).

**X. Were the "BeLaU" spherules concentrated in the presumed path of the bolide?**

Loeb *et al.* claim that the spherules with the distinctive "BeLaU" compositions had a higher concentration in the assumed path of the bolide than out of it. If this were true, this would strongly suggest an association of that composition with an origin in the bolide. However, there is no evidence for variations in the "BeLaU" fraction among collected spherules, and the data are consistent with the "BeLaU" composition being common among background spherules, and uniformly distributed.

Loeb *et al.* collected 622 spherules. Of these, 57 were analyzed: 10 from off the path (runs 17 and 22) and 47 from the path (all other runs). Among these analyzed spherules, 5 were found to have the BeLaU pattern: all 5 were (potentially) on the path, and none were found off the path. It's interesting that they did not find "BeLaU" spherules where they didn't expect to find any, but how hard did they look, really?

The frequency of "BeLaU" spherules among analyzed spherules appears to be 5/57, or 8.8%. If this is characteristic of background spherules, we'd expect the same frequency of "BeLaU" compositions in spherules off the path. However, only 10 such spherules were even analyzed, so we'd expect to find only 0.9, i.e., less than 1 on average. According to Poisson statistics, the probability of finding $k$=0 "BeLaU" spherules among $t$=10 analyzed spherules when the frequency is $r$=0.088, is $(rt)^k \exp(-rt) / k!$ = 42%. That is, there could be just as many "BeLaU"spherules in the control regions, but there's a 42% chance Loeb *et al.* wouldn't find any, because they hardly looked in those regions.

In fact, even if the frequency of "BeLaU" spherules off the path were 17.5%—*double* the frequency on the path—there would *still* be a 17% chance of finding zero "BeLaU" spherules among the 10 analyzed! The typical threshold for significance of a null result is 5%, so the authors cannot even rule out an *increased* abundance of "BeLaU" spherules off the path.

Given the frequency found in spherules on the path, a finding of zero "BeLaU" spherules off the path would not be significant unless at least 34 spherules were analyzed, not just a paltry 10. Only then do the different scenarios predict meaningfully different numbers of "BeLaU" spherules (0 if none are found off the path; 3 if the distribution is uniform everywhere).

> In summary, to do a proper control, Loeb *et al*. should have collected and analyzed many, many more spherules from runs 17 and 22. The data are completely consistent with the "BeLaU" spherules being part of the background of spherules deposited on the seafloor over the last tens of thousands of years, with nothing to do with the 2014-01-08 bolide.

**XI. What is the cause of the enrichments of La, U, etc., in the "BeLaU" spherules?**

Loeb *et al.* speculate considerably about how the chemical compositions of the "BeLaU" spherules could arise. They suggest vaporization played a major role in setting depletions, that siderophile (iron-loving) elements are depleted, and that the most likely explanation is origin from an (extrasolar) planet that had differentiated (melted and separated rock from an iron core). This interpretation has a number of problems. What's striking is that the authors never once considered the chemical effects on the spherules from sitting in sea water.

There is no doubt that vaporization can deplete a spherule in easily-evaporated ('volatile') elements. This probably explains why elements more volatile than Zn *generally* are relatively depleted. It's also clear that volatility is not the only control: Sc is much more depleted than similarly non-volatile ('refractory') Hf and Y; refractory Ca is much more depleted than similarly refractory Ta and Nb; V is not nearly as depleted as similarly volatile Ni and Cr.

They conclude that "This "BeLaU" abundance pattern found in IM1's spherules could have possibly originated from a highly differentiated planetary magma ocean," and "the BeLaU samples possibly reflect an extremely evolved composition from a planetary magma ocean, but not the bodies in the Solar System." It's not clear which part of the planet they mean. The magma ocean would be rich in Mg and highly depleted in Fe, which seems inconsistent with the fact that S21 is < 2wt% Mg and > 60wt% Fe. The rock that crystallizes last from a cooling magma ocean is basalt, but that doesn't seem to be what they mean, either. They compare to Earth's continental crust, which is formed by planetary processes quite distinct from magma oceans. This is the best match, but even so, the "BeLaU" spherules are depleted in Mg and Ca by an order of magnitude relative to continental crust.

Rather than arising in an unusual magma ocean on an extrasolar planet, it is likely that the abundance pattern observed in these spherules is due to the same process that gave rise to the abundance patterns in the Indian Ocean spherules. The sediments in which the Indian Ocean spherules were found were also measured by Prasad et al. (2015) for their elemental compositions and found to be enriched in these same elements (La, U, Th, etc.). Rudraswami et al. (2016) interpreted their spherules to have chemically interacted with these sediments (mediated by seawater) over the ~50 kyr residence time at the seafloor. Likewise, van Ginneken et al. (2021) attributed enrichments in La, Th, and U in their micrometeorites to terrestrial contamination. Given the low sedimentation rates in the site where the authors recovered their samples, it is to be expected that they had a similarly long residence time and chemically interacted with their surrounding sediments. A much greater loss of Mg and Ca is the only discernible difference between these samples and the Indian Ocean spherules, and is easily explained by longer residence in the ocean.

> In summary, the "BeLaU" abundance patterns in the Papua New Guinea spherules probably have the same origin as the Indian Ocean spherules: water-driven chemical reactions with the sea water and the seafloor sediments the spherules are in. The "BeLaU" spherules like S21 probably sat on the seafloor for tens of thousands of years.

**XII. Does the Be in the spherules indicate GCR irradiation in interstellar space?**

Loeb *et al.* attach significance to the abundance of Be in their 5 "BeLaU" spherules. The mass fraction of Be in them averages about 300 times the concentration in CI chondrites, or ~8 ppm. They claim this abundance is not only consistent with, but an *indicator* of creation of Be by Galactic cosmic rays (GCRs) striking nuclei in the sample over billions of years while traveling through interstellar space: "… the enhanced Be abundance might represent a flag of cosmic-ray spallation on IM1's surface along a [sic] extended interstellar journey… this constitutes a fourth indicator of an interstellar origin to IM1…"

In principle, billions of years of exposure to GCRs would convert a fraction of the oxygen nuclei into Li, Be, and B nuclei; but the fraction is very small. After millions of years of GCR exposure, meteorites acquire $^9$Be and $^{10}$Be abundances on the order of $10^{-6}$ ppm (equivalent to tens of disintegrations per minute of radioactive $^{10}$Be). This is much, much smaller than the ~0.03 ppm abundances of Be meteorites typically have due to normal chemical processes. This is why only radioactive $^{10}$Be, not overall Be itself, is the flag of GCR exposure After all, gems of emerald and ruby and aquamarine on Earth are 5% beryllium, but this is not an indicator of an interstellar origin. After billions of years, the GCR-produced Be abundance might be as large as ~$10^{-3}$ ppm, as corroborated by a simple calculation [assuming a flux of > 30 MeV GCRs $F$ ~ 10 cm$^{-2}$ s$^{-1}$, a cross section for Be production of σ ~ 6 mbarns, and an exposure time $t$ ~ 1 Gyr, a fraction ($F σ t$) ~ 2 x $10^{-9}$ of the oxygen nuclei would be converted into $^9$Be and $^{10}$Be nuclei, yielding a mass fraction of Be ~ 0.0006 ppm]. But even exposure to GCRs for billions of years would not produce even one part in $10^4$ of the Be measured in the "BeLaU" cosmic spherules.

The enhanced (above CI chondrite levels) abundances of Be in the spherules clearly is due to chemistry. There is ample opportunity for to add Be to the cosmic spherules via aqueous alteration while sitting on the seafloor for tens of thousands of years. Deep ocean water worldwide contains Be at an abundance of 20-30 pM (picomolar), supplied by river particulates (von Blanckenburg et al. 1996). Be atoms precipitate into ferromanganese nodules after spending < 1 kyr in the oceans, yielding Be concentrations in the nodules of 2-3 ppm (Merrill et al. 1960; von Blanckenburg et al. 1996). Cosmic spherules exposed to sea water would have comparable Be concentrations, consistent with the ~8 ppm measured in the "BeLaU" spherules.

> In summary, even billions of years of GCR irradiation wouldn't produce even a tiny fraction of the Be seen in the "BeLaU" spherules. It must be due to chemistry like that experienced by nodules on the seafloor, and indeed the Be concentration in the spherules is comparable to the Be concentration in ferromanganese nodules.

**XIII. Conclusions**

Loeb *et al.* claim the spherules they found are "likely" to be interstellar in origin. They advanced arguments to this effect, but upon slight scrutiny, all of their arguments fall apart. There is no evidence for interstellar materials. The 2014-01-08 probably wasn't interstellar. If it were, it would have completely vaporized. Even if ablation spherules were produced, these would have been few and spread out, and vastly outnumbered by background spherules. There is no evidence that spherules overall, or "BeLaU" spherules, were concentrated anywhere, let alone the path of the bolide, which is very poorly known. The "BeLaU" triple spherule S21 most likely formed in an impact plume tens of thousands of years ago, and not in a fireball like the 2014-01-08 bolide. The "BeLaU" pattern is seen in other cosmic spherules in the Indian Ocean and Antarctica and is attributed to terrestrial contamination. The Be abundance also is attributable to terrestrial contamination, not cosmic-ray spallation. Finally, the Fe isotopes are more consistent with terrestrial contamination than vaporization during entry, and are a smoking gun for a Solar System origin for all their spherules. Not a single one of their arguments holds water.

The reason their arguments fall apart so easily is because they did not follow the scientific method. They made associations between their data and their favored hypothesis (interstellar origin), but at no point did they consider any competing alternative hypothesis and ask whether the data are better explained by that model. It's a textbook example of **confirmation bias**.

**The simplest competing hypothesis is this: the spherules they collected are part of the copious background of cosmic spherules deposited on the seafloor over the world.**

During the tens of thousands of years spherules spend on the seafloor before being buried by sediments, they may chemically react with the sea water and the sediments. This alternative hypothesis could have been tested by sampling the sediments along with the spherules, or by collecting a large and statistically meaningful number of spherules from regions far from the bolide's path. Because they didn't consider alternative hypotheses, Loeb *et al.* did not properly design their experiment to avoid inconclusive results.

On January 31, 2021, Loeb is quoted by *The Guardian*[17] as saying: "If someone comes to me and says, 'For these scientific reasons, I have a scenario that makes much more sense than yours,' then I'd rip that paper up and accept it," he says. "But most of the people who attacked, they hadn't even looked at my paper, or read the issues, or referred to the items we discussed…"

We have read and considered the manuscript they uploaded to arXiv, and we have a scenario that makes much more sense than theirs. Will they now metaphorically rip up their paper?

---

[17] https://www.theguardian.com/science/2021/jan/31/professor-avi-loeb-it-would-be-arrogant-to-think-were-alone-in-the-universe-

Blanchard, M. B., Brownlee, D. E., Bunch, T. E., Hodge, P. W., and Kyte, F. T. (1980) "Meteoroid ablation spheres from deep-sea sediments" *Earth & Planet. Sci. Lett*. 46, 178-190.

Borovicka, J., Spurny, P., Grigore, V. I., and Svoren, J. (2017) "The January 7, 2015 superbolide over Romania and structural diversity of meter-sized asteroids", *Planet. & Space Sci.* 143, 147-158.

Brown, P. G., Wiegert, P., Clark, D. and Tagliaferri, E. (2016) "Orbital and physical characteristics of meter-scale impactors from airburst observations", *Icarus* 266, 96-111.

Brown, P. G. and Borovicka, J. (2023) "On the Proposed Interstellar Origin of the USG 20140108 Fireball", *Ap.J.* 953, 167-179.

Brownlee, D. E. (1985) "Cosmic dust – Collection and research", *Ann. Rev. Earth & Planet. Sci.* 13, 147-173.

Carrillo, A., Ness, M. K., Hawkins, K., Sanderson, R. E., Wang, K., Wetzel, A., and Bellardini, M. A. (2023) "The Relationship between Age, Metallicity, and Abundances for Disk Stars in a Simulated Milky Way", *Ap. J.* 942, 35-56.

Desch, S. J., Jackson, A. P. and Hartnett, H. E. (2023), "The Challenges of Recovering Interstellar Meteorites", Asteroids, Comets and Meteors Conference 14, #2244 (abstract).

Devillepoix, H. A. R., Bland, P. A., Sansom, E. K., Towner, M. C., Cupak, M., Howie, R. M., Hartig, B. A. D., Jansen-Sturgeon, T. & Cox, M. A. (2019) "Observations of metre-scale impactors by the Desert Fireball Network", *Mon. Not. Roy. Astron. Soc.* 483, 5166-5178.

Doyle, L. J., Hopkins, T. L. and Betzer, P. R. (1976) "Black magnetic spherule fallout in the eastern Gulf of Mexico", *Science* 194, 1157-1159.

Folco, L., and Cordier, C. (2015) "Micrometeorites" in *Planetary Mineralogy*, M. R. Lee and H. Leroux, eds., EMU Notes in Mineralogy 15, 253-297.

Gallardo, P. A. (2023) "Anthropogenic coal ash as a contaminant in a micro-meteoritic underwater search", *Res. Notes. Amer. Astron. Soc.* 7, 220.

Kobayashi, C., Karakas, A. I., and Umeda, H. (2011) "The evolution of isotope ratios in the Milky Way Galaxy", *Mon. Not. Roy. Astron. Soc.* 414, 3231-3250.

Lodders, K. (2003) "Solar System Abundances and Condensation Temperatures of the Elements", *Ap.J*. 591, 1220-1247.

Merrill, J. R., Lyden, E. F. X., Honda, M., and Arnold, J. R. (1960) "The sedimentary geochemistry of the beryllium isotopes", *Geochim. Cosmochim. Acta* 18, 108-129.

Millard, H. T., Jr., and Finkelman, R. B. (1970) "Chemical and mineralogical compositions of cosmic and terrestrial spherules from a marine sediment", *J. Geophys. Res*. 75, 2125-2134.

Milley, E. P. (2010) "Physical Properties of Fireball-Producing Earth-Impacting Meteoroids and Orbit Determination through Shadow Calibration of the Buzzard Coulee Meteorite Fall", Ph.D. Thesis (University of Calgary).

Moilanen, J., Gritsevich, M. and Lyytinen, E. (2021) "Determination of strewn fields for meteorite falls", *Mon. Not. Roy. Astron. Soc.* 503, 3337-3350.

Petrovic, J. J. (2001) "Mechanical properties of meteorites and their constituents", *J. Mat. Sci.* 36, 1579-1583.

Pohl, L. and Britt, D. T. (2020) "Strengths of meteorites—An overview and analysis of available data", *Meteoritics & Plan. Sci.* 55, 962-987.

Popova, O., Borovicka, J., and Campbell-Brown, M. D. (2019) "Modeling the Entry of Meteoroids", in *Meteoroids: Sources of Meteors on Earth and Beyond,* G. O. Ryabova, D. J. Asher, and M. D. Campbell-Brown, eds. (Cambridge Univ. Press), pp. 9-36.

Prasad, M. Shyam, Rudraswami, N. G., de Araujo, A., Babu, E. V. S. S. K., and Vijaya Kumar, T. (2015) "Ordinary chondritic micrometeorites from the Indian Ocean", *Meteoritics & Planet. Sci.* 50, 1013-1031.

Revelle, D. O. and Ceplecha, Z. (1994) "Analysis of identified iron meteoroids: possible relation with M-type Earth-crossing asteroids?", *Astron. & Astrophys.* 292, 330-336.

Revelle, D. O. and Ceplecha, Z. (2001) "Bolide physical theory with application ot PN and EN fireballs", in *Proceedings of the Meteoroids 2001 Conference* (ESA SP-2495), 507-512.

Rudraswami, N. G., Parashar, K., and Shaym Prasad, M. (2011) "Micrometer- and nanometer-sized platinum group nuggets in micrometeorites from deep-sea sediments of the Indian Ocean", *Meteoritics. & Planet. Sci.* 46, 470-491.

Rudraswami, N. G., Shaym Prasad, M., Babu, E. V. S. S. K., and Vijaya Kumar, T. (2016) "Major and trace element geochemistry of S-type cosmic spherules", *Meteoritics & Planet. Sci.* 51, 718-742.

Siraj, A. & Loeb, A. (2022a) "A Meteor of Apparent Interstellar Origin in the CNEOS Fireball Catalog", *Ap.J.* 939, 53-57.

Siraj, A. & Loeb, A. (2022b) "Interstellar Meteors are Outliers in Material Strength", *Ap.J.L.* 941, L28-31.

Siraj, A. & Loeb, A. (2023) "Localizing the First Interstellar Meteor with Seismometer Data", arXiv:2303.07357 (unpublished).

van Ginneken, M., Goderis, S., Artemieva, N., Debaille, V., Decree, S., Harvey, R. P., Huwig, K. A., Hecht, L., Yang, S., Kaufmann, F. E. D., Soens, B., Humayun, M., van Maldeghem, Genge, M. J., and Claeys, P. (2021) "A large meteoritic event over Antarctica ca. 430 ka ago inferred from chondritic spherules from the Sor Rondane Mountains", *Sci. Adv.* 7, eabc1008.

von Blanckenburg, F., O'Nions, R. K., Belshaw, N. S., Gibb, A., and Hein, J. R. (1996), "Global distribution of beryllium isotopes in deep ocean water as derived from Fe-Mn crusts", *Earth & Planet. Sci. Lett.* 141, 213-226.

Zuluaga, J. I. (2019) "Speed Thresholds for Hyperbolic Meteors: The Case of the 2014 January 08 CNEOS Meteor", *Res. Notes Amer. Astron. Soc.* 3, 68.